\documentclass[a4paper]{jpconf}
\usepackage{graphicx}
\usepackage{subfigure}
\begin{document}
\title{A Study of the QCD Critical Point Using Particle Ratio Fluctuations}

\author{Terence J Tarnowsky (for the STAR Collaboration)}

\address{National Superconducting Cyclotron Laboratory,
Michigan State University, East Lansing, MI 48824, USA}

\ead{tarnowsk@nscl.msu.edu}

\begin{abstract}
Dynamical fluctuations in global conserved quantities such as baryon number, strangeness, or charge may be observed near a QCD critical point. Results from new measurements of dynamical $K/\pi$ and $p/\pi$ ratio fluctuations are presented. The commencing of a QCD critical point search at RHIC has extended the reach of possible measurements of dynamical $K/\pi$ and $p/\pi$ ratio fluctuations from Au+Au collisions to lower energies. The STAR experiment has performed a comprehensive study of the energy dependence of these dynamical fluctuations in Au+Au collisions at the energies $\sqrt{s_{NN}}$ = 7.7, 11.5, and 39 GeV. New results are compared to previous measurements and to theoretical predictions from several models.
\end{abstract}

\section{Introduction}
Fluctuations and correlations are well known signatures of phase transitions. In particular, the quark/gluon to hadronic phase transition may lead to significant fluctuations \cite{Koch1}. In 2010, the Relativistic Heavy Ion Collider (RHIC) began a program to search for the QCD critical point. This involves an ``energy scan'' of Au+Au collisions from top collision energy ($\sqrt{s_{NN}}$ = 200 GeV) down to energies as low as $\sqrt{s_{NN}}$ = 7.7 GeV \cite{STARBES}. This critical point search will make use of the study of correlations and fluctuations, particularly those that could be enhanced during a phase transition that passes close to the critical point \cite{kurtosis}. Particle ratio fluctuations are an observable that has already been studied as a function of energy and system-size. Combining these previous measurements with new results from the energy scan will provide additional information to assist in locating the QCD critical point.

Dynamical particle ratio fluctuations, specifically fluctuations in the $K/\pi$ and $p/\pi$ ratio, can provide information on the quark-gluon to hadron phase transition \cite{Strangeness1, Strangeness2, Strangeness3}. The variable used to measure these fluctuations is referred to as $\nu_{dyn}$. $\nu_{dyn}$ was originally introduced to study net charge fluctuations \cite{nudyn1, nudyn2}. $\nu_{dyn}$ quantifies deviations in the particle ratios from those expected for an ideal statistical Poissonian distribution. The definition of $\nu_{dyn,K/\pi}$ (describing fluctuations in the $K/\pi$ ratio) is,
\begin{eqnarray}
\nu_{dyn,K/\pi} = \frac{<N_{K}(N_{K}-1)>}{<N_{K}>^{2}}
+ \frac{<N_{\pi}(N_{\pi}-1)>}{<N_{\pi}>^{2}}
- 2\frac{<N_{K}N_{\pi}>}{<N_{K}><N_{\pi}>}\ ,
\label{nudyn}
\end{eqnarray}
where $N_{K}$ and $N_{\pi}$ are the number of kaons and pions in a particular event, respectively. In this proceeding, $N_{K}$ and $N_{\pi}$ are the total charged multiplicity for each particle species. A formula similar to (\ref{nudyn}) can be constructed for $p/\pi$ and other particle ratio fluctuations. By definition, $\nu_{dyn}$ = 0 for the case of a Poisson distribution of kaons and pions. It is also largely independent of detector acceptance and efficiency in the region of phase space being considered \cite{nudyn2}. An in-depth study of $K/\pi$ fluctuations in Au+Au collisions at $\sqrt{s_{NN}}$ = 200 and 62.4 GeV was previously carried out by the STAR experiment \cite{starkpiprl}.

Earlier measurements of particle ratio fluctuations utilized the variable $\sigma_{dyn}$ \cite{NA49}, 
\begin{equation}
\sigma_{dyn} = sgn(\sigma_{data}^{2}-\sigma_{mixed}^{2})\sqrt{|\sigma_{data}^{2}-\sigma_{mixed}^{2}|}\ ,
\label{signudyn}
\end{equation}
where $\sigma$ is the relative width of the $K/\pi$ or $p/\pi$ distribution in either real data or mixed events. It has been shown that $\nu_{dyn}$ is a first order expansion of $\sigma_{dyn}$ about the mean of its denominator \cite{jeon,baym,sdasthesis}. The two variables are related as $\sigma_{dyn}^{2} \approx \nu_{dyn}$.

\section{Experimental Analysis}
The data presented here for $K/\pi$ and $p/\pi$ fluctuations was acquired by the STAR experiment at RHIC from minimum bias (MB) Au+Au collisions at center-of-mass collision energies ($\sqrt{s_{NN}}$) of 200, 130, 62.4, 39, 19.6, 11.5, and 7.7 GeV \cite{STAR}. The main particle tracking detector at STAR is the Time Projection Chamber (TPC) \cite{STARTPC}. All detected charged particles in the pseudorapidity interval $|\eta| < 1.0$ were measured. The transverse momentum ($p_{T}$) range for pions and kaons was $0.2 < p_{T} < 0.6$ GeV/$c$, and for protons was $0.4 < p_{T} < 1.0$ GeV/$c$. Charged particle identification involved measured ionization energy loss ($dE/dx$) in the TPC gas and total momentum ($p$) of the track. The energy loss of the identified particle was required to be less than two standard deviations (2$\sigma$) from the predicted energy loss of that particle at a particular momentum. A second requirement was that the measured energy loss of a pion/kaon was more than 2$\sigma$ from the energy loss prediction of a kaon/pion, and similarly for $p/\pi$ measurements. The centralities used in this analysis account for 0-10, 10-20, 20-30, 30-40, 40-50, 50-60, 60-70, and 70-80\% of the total hadronic cross section.

Data was also analyzed using the recently completed Time of Flight (TOF) detector \cite{STARTOF}. The TOF is a multi-gap resistive plate chamber (MRPC) detector. Particle identification was carried out using the time-of-flight information for a track along with its momentum, determined by the TPC. From these two quantities the particle's mass can be calculated. Utilizing the TOF allows the identified particle momentum reach to be extended. TOF identification extends the $p_{T}$ range of pions and kaons from, $0.6 < p_{T} < 1.4$ GeV/$c$ and protons from $1.0 < p_{T} < 1.8$ GeV/$c$.

\section{Results and Discussion}

The STAR experiment has recently acquired data at three new energies. The first phase of the QCD critical point search involved Au+Au collisions at $\sqrt{s_{NN}}$ = 39, 11.5, and 7.7 GeV. STAR has measured dynamical $K/\pi$ and $p/\pi$ fluctuations at all these energies.

\begin{figure}
\subfigure[]{
\includegraphics[width=0.55\textwidth]{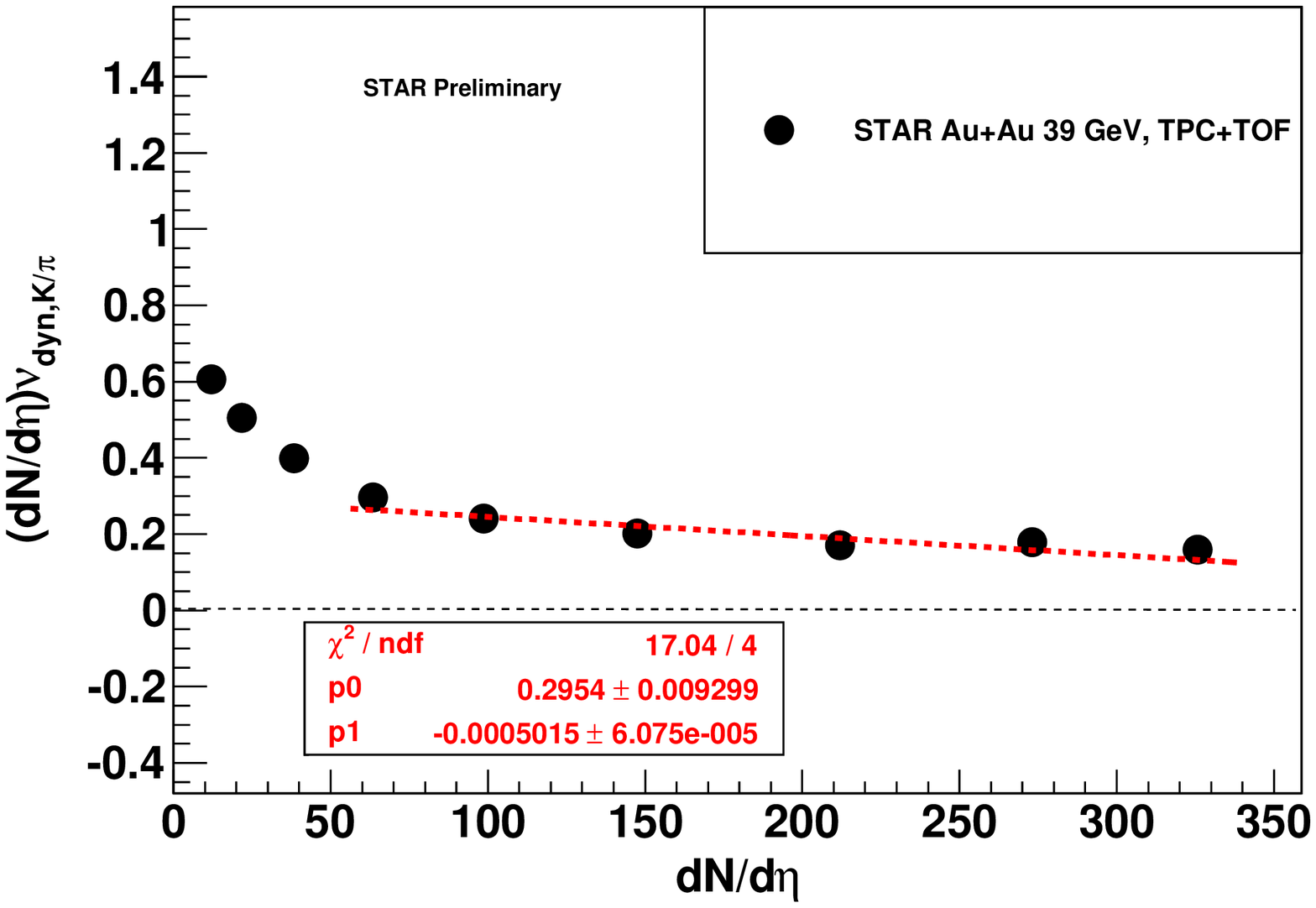} 
\label{kpi_39gev}
}
\subfigure[]{
\includegraphics[width=0.55\textwidth]{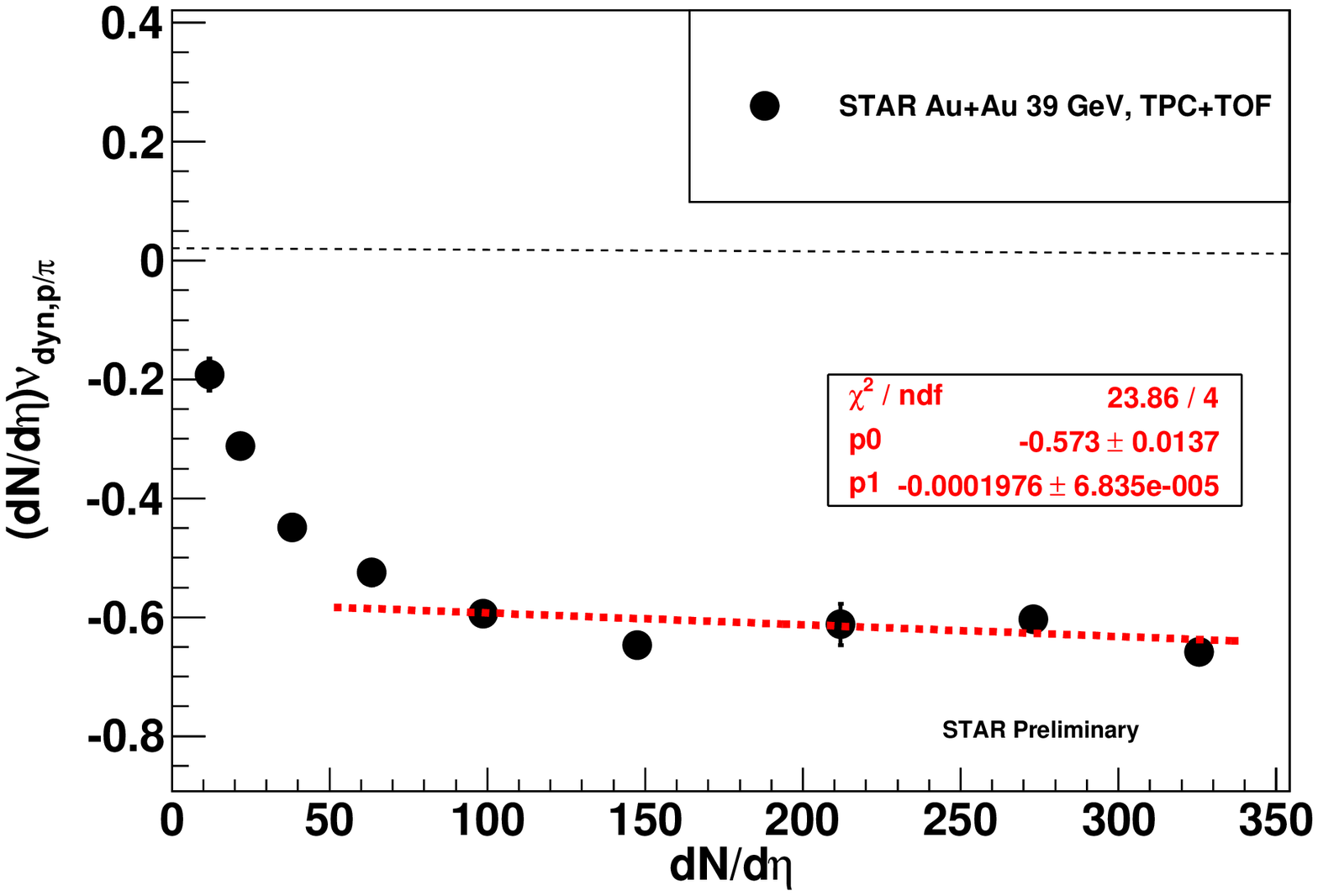} 
\label{ppi_39gev}
}
\caption{Results for the measurement of $\nu_{dyn,K/\pi}$ (a) and $\nu_{dyn,p/\pi}$ (b) scaled by uncorrected charged particle multiplicity, $dN/d\eta$, as measured by the STAR TPC+TOF (black stars) from central 0-5\% Au+Au collisions at $\sqrt{s_{NN}}$ = 39 GeV. Errors shown are statistical.}
\label{asdf}
\end{figure}

Figure \ref{kpi_39gev} shows the measurement of $\nu_{dyn}$ for $K/\pi$ fluctuations as a function of centrality, scaled by the uncorrected charged particle multiplicity, $dN/d\eta$, from Au+Au collisions at $\sqrt{s_{NN}}$ = 39 GeV. The measured $dN/d\eta$ is a raw value, not corrected for efficiency and acceptance. Scaling by $dN/d\eta$ removes the 1/$N_{ch}$ dependence of $\nu_{dyn}$ \cite{nudyn1}, where $N_{ch}$ is the charged particle multiplicity. The data points include charged particles identified using TPC $dE/dx$ for low momenta and mass-squared (m$^{2}$) by the STAR TOF for higher momenta. The errors shown are statistical. Dynamical $K/\pi$ fluctuations are always positive. In peripheral collisions at higher energies ($\sqrt{s_{NN}}$ = 62.4 and 200 GeV), $\nu_{dyn,K/\pi}$ was found to scale linearly with $dN/d\eta$. A linear fit to the data points and its parameters is shown for $dN/d\eta > $ 50.  As seen in Figure \ref{kpi_39gev}, this linear scaling is broken in very peripheral collisions for $dN/d\eta$ less than 50 charged particles. 

Figure \ref{ppi_39gev} shows the measurement of $\nu_{dyn,p/\pi}$ as a function of centrality, scaled by the uncorrected charged particle multiplicity, $dN/d\eta$, from Au+Au collisions at $\sqrt{s_{NN}}$ = 39 GeV. A linear fit to the data for $dN/d\eta > $ 50 is also included. Dynamical $p/\pi$ fluctuations are always negative due to correlated production of protons and pions from resonances such as the $\Delta$. Similar to $\nu_{dyn,K/\pi}$ at $\sqrt{s_{NN}}$ = 39 GeV, dynamical $p/\pi$ fluctuations are approximately centrality independent until $dN/d\eta < $ 50, where they become less negative in the peripheral bins. 

Accounting for track matching efficiency and detector acceptance, the TOF detection efficiency is approximately $\frac{2}{3}$ that of the TPC. Differences between the two data samples are still under investigation.

\begin{figure}
\subfigure[]{
\includegraphics[width=0.55\textwidth]{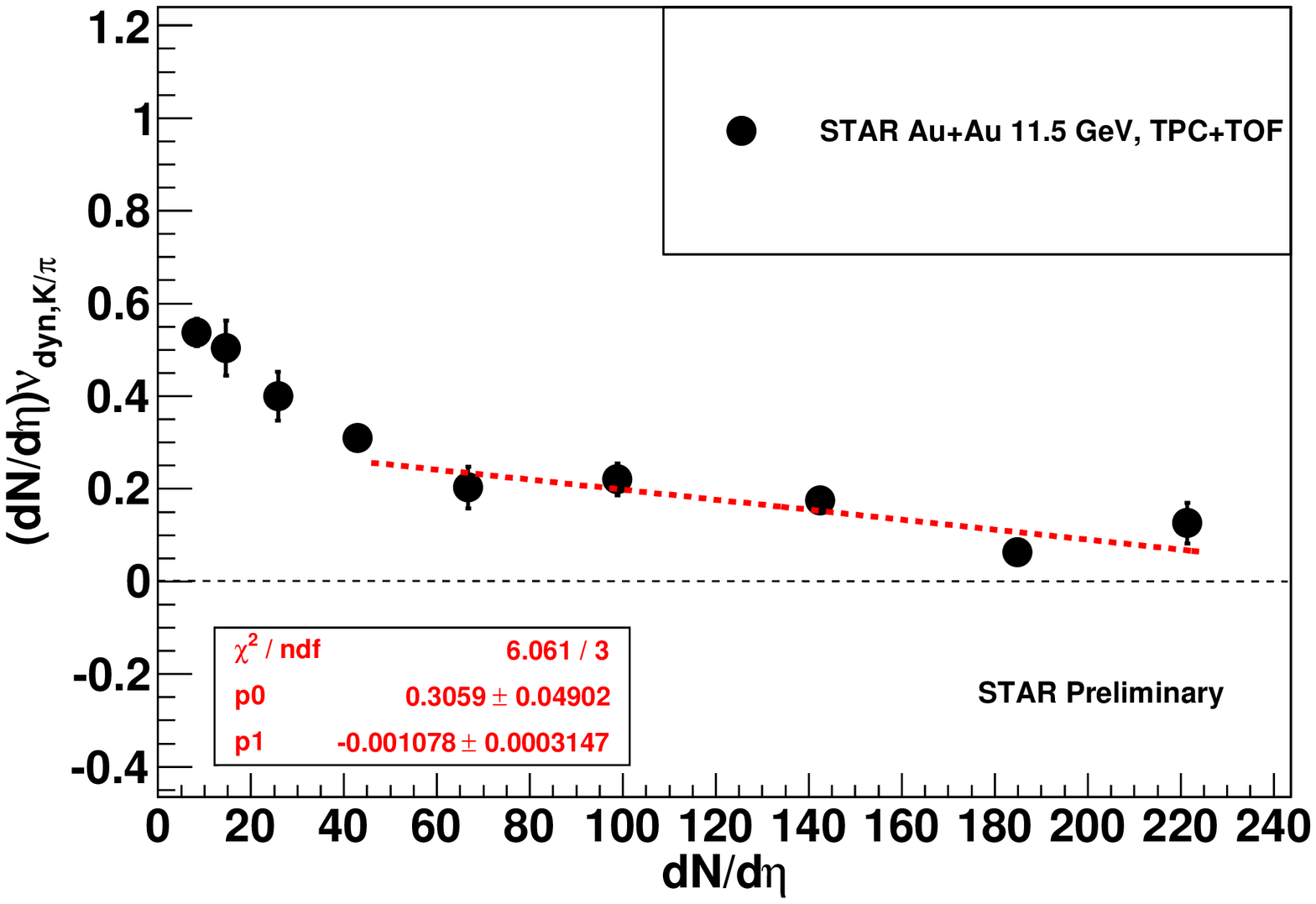} 
\label{kpi_11gev}
}
\subfigure[]{
\includegraphics[width=0.55\textwidth]{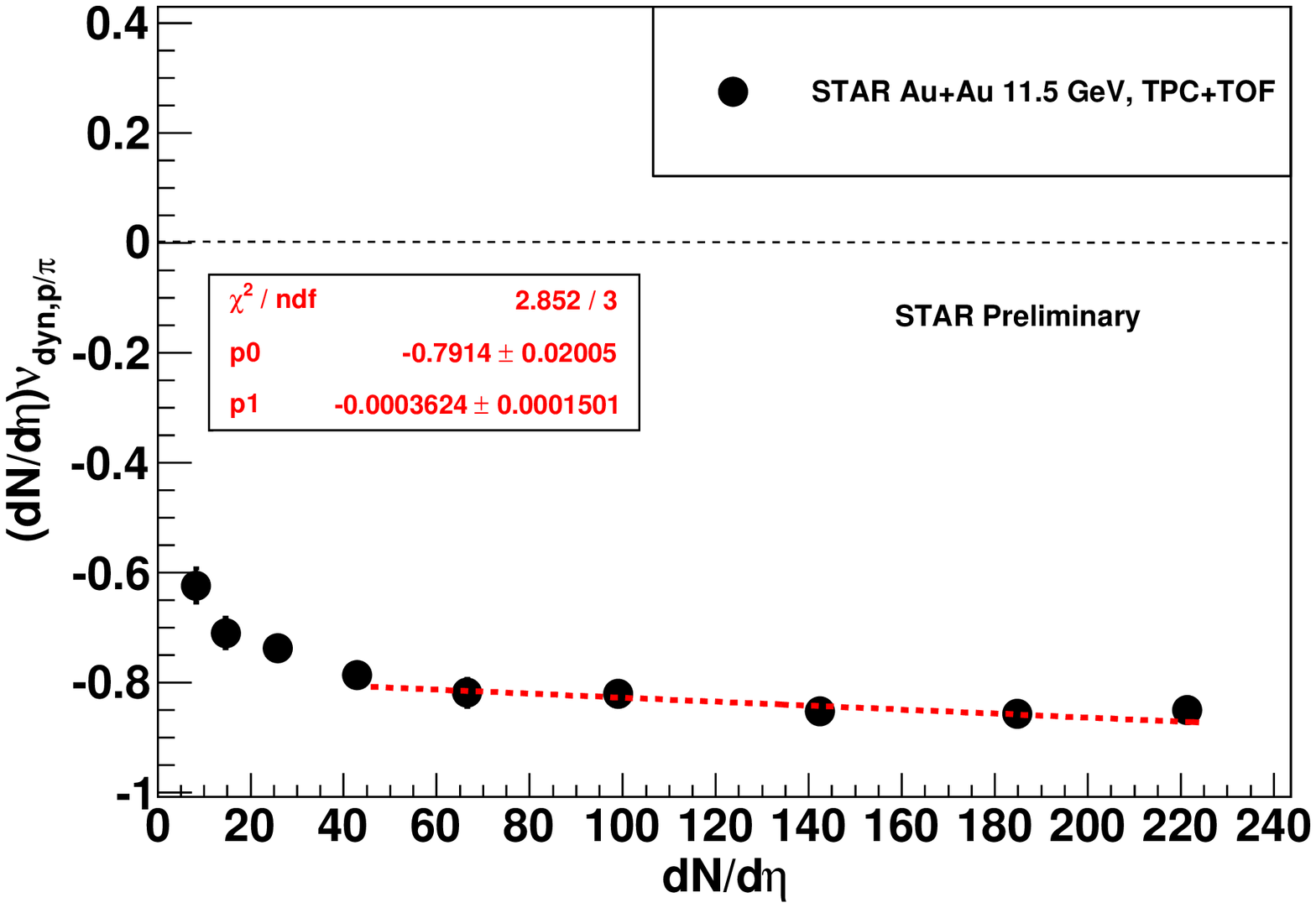} 
\label{ppi_11gev}
}
\caption{Results for the measurement of $\nu_{dyn,K/\pi}$ (a) and $\nu_{dyn,p/\pi}$ (b) scaled by uncorrected charged particle multiplicity, $dN/d\eta$, as measured by the STAR TPC+TOF (black circles) from central 0-5\% Au+Au collisions at $\sqrt{s_{NN}}$ = 11.5 GeV. Errors shown are statistical.}
\label{asdf}
\end{figure}

Figure \ref{kpi_11gev} shows the measurement of $\nu_{dyn,K/\pi}$ as a function of centrality, scaled by the uncorrected charged particle multiplicity, $dN/d\eta$, from Au+Au collisions at $\sqrt{s_{NN}}$ = 11.5 GeV. The data points include charged particles identified using TPC $dE/dx$ for low momenta and mass$^{2}$ by the STAR TOF for higher momenta. The errors shown are statistical. A linear fit to the data points and its parameters is shown for $dN/d\eta > $ 40. In these early results, there is no apparent scaling with $dN/d\eta$ for $\nu_{dyn,K/\pi}$ at $\sqrt{s_{NN}}$ = 11.5 GeV. The magnitude of the dynamical $K/\pi$ fluctuations at $\sqrt{s_{NN}}$ = 11.5 GeV have a similar range from peripheral to central collisions as those shown in Figure \ref{kpi_39gev} at $\sqrt{s_{NN}}$ = 39 GeV.

Figure \ref{ppi_11gev} shows the measurement of $\nu_{dyn,p/\pi}$ as a function of centrality, scaled by the uncorrected charged particle multiplicity, $dN/d\eta$, from Au+Au collisions at $\sqrt{s_{NN}}$ = 11.5 GeV. A linear fit to the data for $dN/d\eta > $ 40 is also included. The multiplicity scaled $\nu_{dyn,p/\pi}$ at $\sqrt{s_{NN}}$ = 11.5 GeV does not show a rapid change as a function of centrality in peripheral collisions, as the same observable does at $\sqrt{s_{NN}}$ = 39 GeV in Figure \ref{ppi_39gev}. In central collisions, the multiplicity scaled $\nu_{dyn,p/\pi}$ at $\sqrt{s_{NN}}$ = 11.5 GeV measures $\approx$ -0.8, whereas in central collisions at $\sqrt{s_{NN}}$ = 39 GeV it reaches a value of $\approx$ -0.6. 

\begin{figure}
\subfigure[]{
\includegraphics[width=0.55\textwidth]{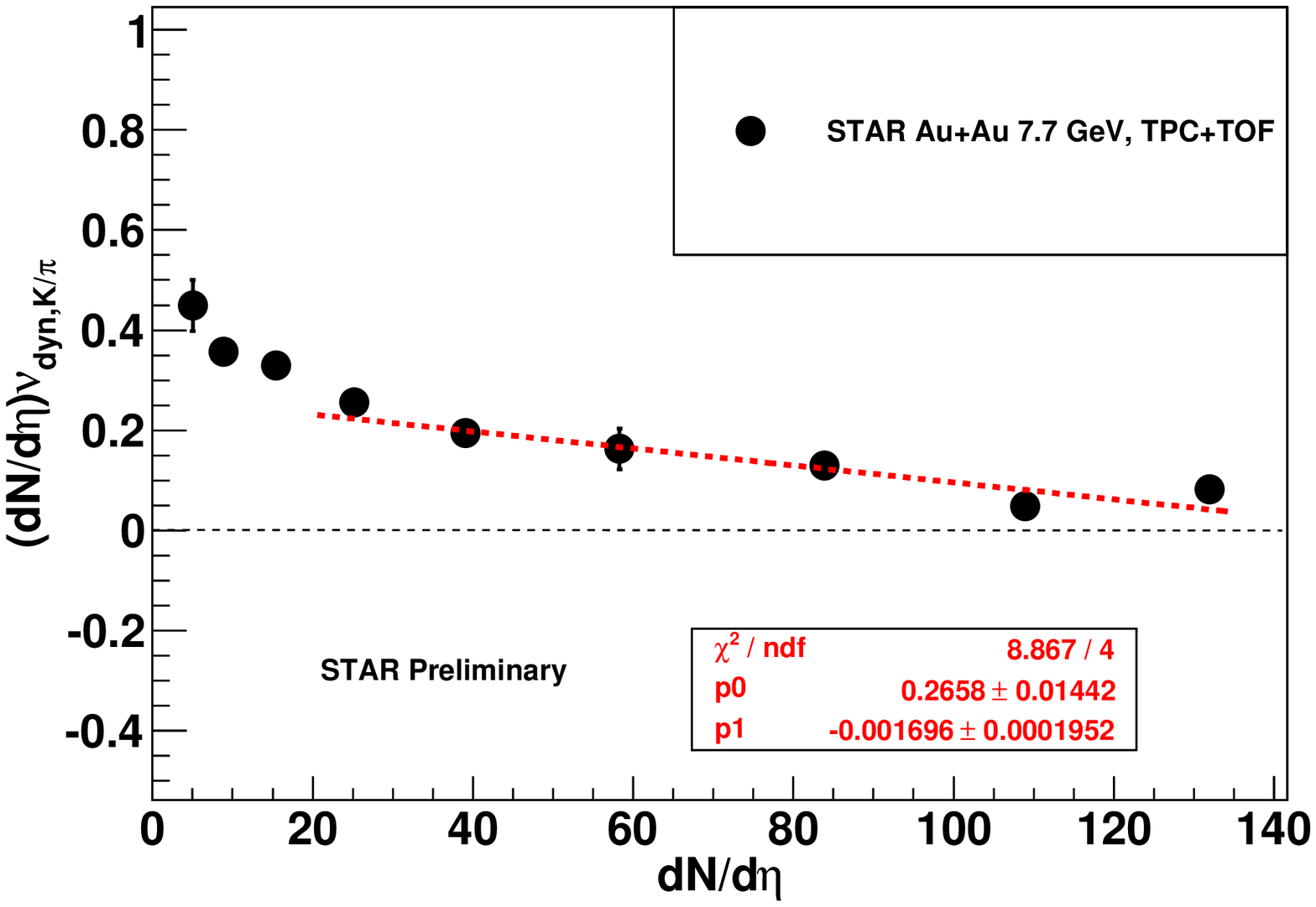} 
\label{kpi_7gev}
}
\subfigure[]{
\includegraphics[width=0.55\textwidth]{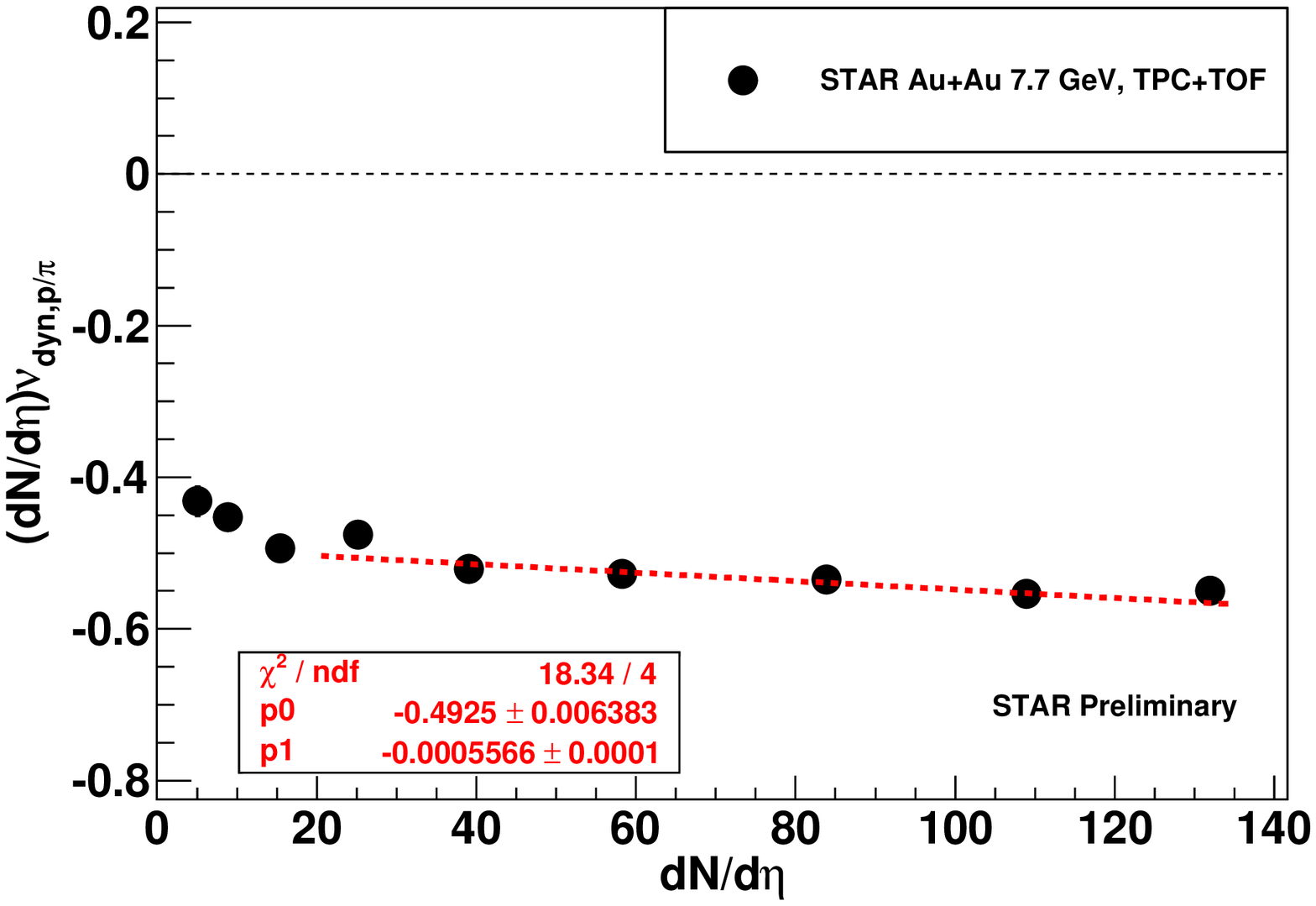} 
\label{ppi_7gev}
}
\caption{Results for the measurement of $\nu_{dyn,K/\pi}$ (a) and $\nu_{dyn,p/\pi}$ (b) scaled by uncorrected charged particle multiplicity, $dN/d\eta$, as measured by the STAR TPC+TOF (black circles) from central 0-5\% Au+Au collisions at $\sqrt{s_{NN}}$ = 7.7 GeV. A linear fit to the TPC data and its parameters are also shown. Errors shown are statistical.}
\label{asdf}
\end{figure}

Figure \ref{ppi_7gev} includes the measurement of dynamical $p/\pi$ fluctuations in Au+Au collisions at $\sqrt{s_{NN}}$ = 7.7 GeV. This is currently the lowest energy data available at RHIC. As with dynamical fluctuations at $\sqrt{s_{NN}}$ = 11.5 and 39 GeV, $\nu_{dyn,p/\pi}$ is plotted as a function of centrality and scaled by the uncorrected charged particle multiplicity, $dN/d\eta$. At this low energy, the antibaryon/baryon ratio is $\sim$ 0.01 \cite{NA49_pbarp}. Thus, the measurement of $p/\pi$ fluctuations at this energy is dominated by protons. Measurements using both TPC and TOF particle identification are strongly negative and approximately independent of centrality. This can be compared to Figures \ref{ppi_39gev} and \ref{ppi_11gev} where the scaled $\nu_{dyn,p/\pi}$ exhibits a centrality dependence in peripheral collisions. In central collisions, it reaches a magnitude of $\approx$ -0.6, which is the same value seen at $\sqrt{s_{NN}}$ = 39 GeV. The dynamics of particle production change with energy and additional studies of charged species dependence may provide information on the behavior of these scalings.

\begin{figure}
\subfigure[]{
\includegraphics[width=0.55\textwidth]{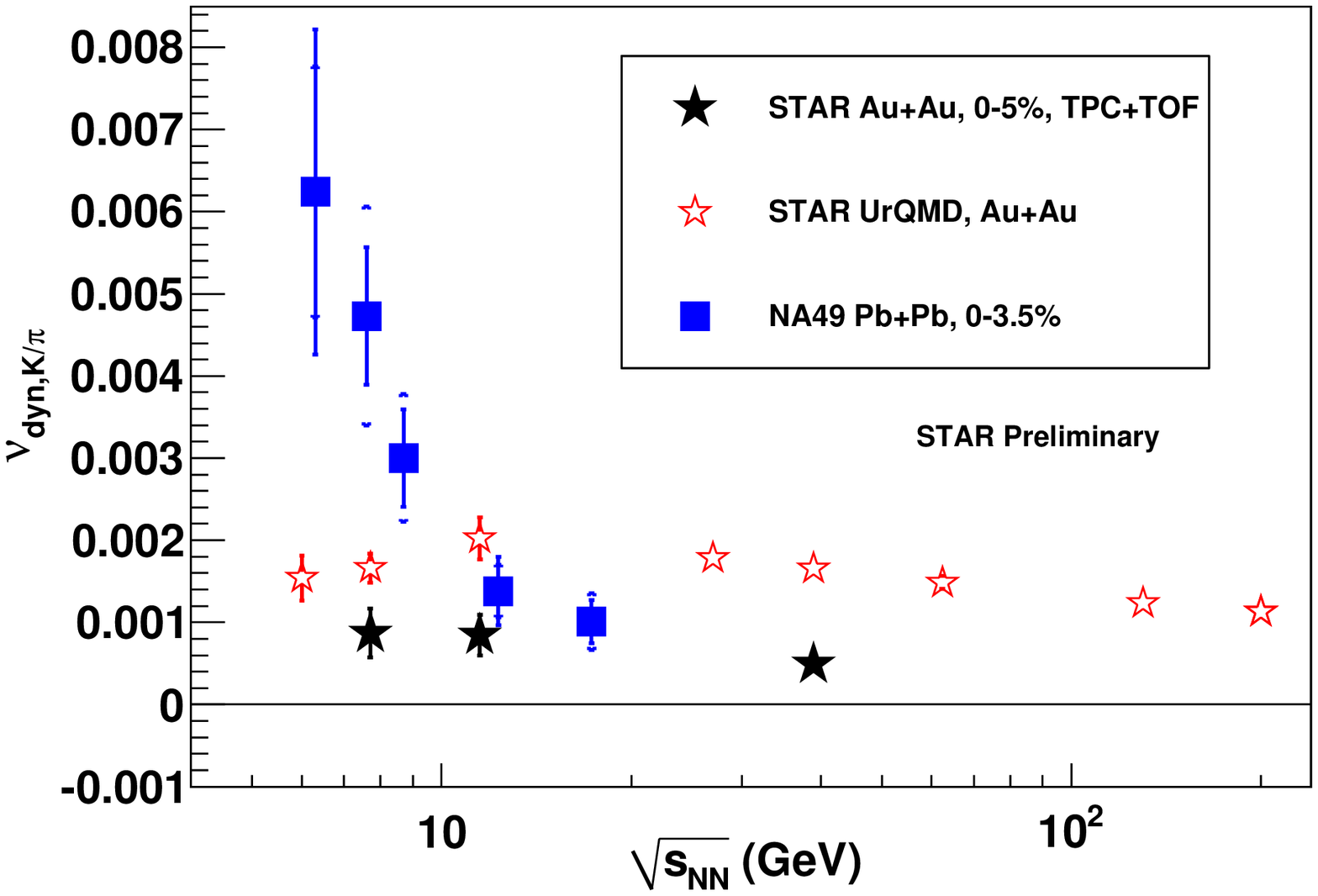} 
\label{kpi_excitation}
}
\subfigure[]{
\includegraphics[width=0.55\textwidth]{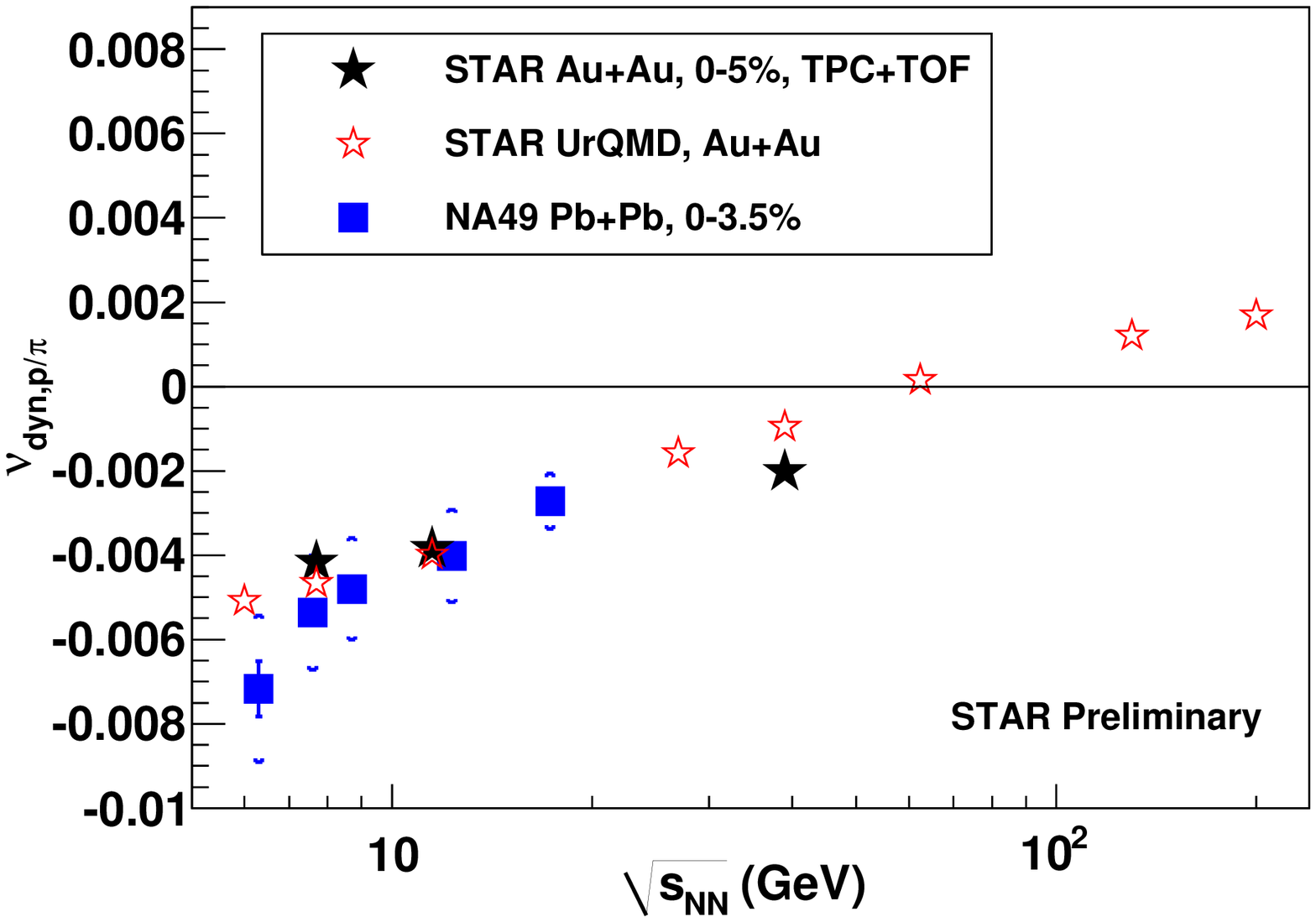} 
\label{ppi_excitation}
}

\caption{Results for the measurement of $\nu_{dyn,K/\pi}$ (a) and $\nu_{dyn,p/\pi}$ (b) as measured by the STAR TPC+TOF (black stars) from central 0-5\% Au+Au collisions at $\sqrt{s_{NN}}$ = 7.7-39 GeV.  Also shown are the same measurements from the NA49 experiment (solid blue squares) from central 0-3.5\% Pb+Pb collisions. Model predictions from UrQMD using the STAR experimental acceptance (red open stars)  are also included.}
\label{asdf}
\end{figure}

Figure \ref{kpi_excitation} shows the measured dynamical $K/\pi$ fluctuations as a function of incident energy, expressed as $\nu_{dyn,K/\pi}$. The measured fluctuations from the STAR experiment are plotted as the black stars, measurements from the NA49 experiment \cite{NA49_kpi_ppi} as blue squares, a transport model prediction from UrQMD (red open stars), using the STAR experimental acceptance. The published NA49 results have been converted to $\nu_{dyn}$ using the relation $\sigma_{dyn}^{2} \approx \nu_{dyn}$. STAR measures dynamical $K/\pi$ fluct`uations that are approximately independent of collision energy from $\sqrt{s_{NN}}$ = 7.7-39 GeV. The NA49 experiment observes a strong decrease with increasing incident energy for central 0-3.5\% Pb+Pb collisions (solid blue squares). Particle production mechanisms change with energy, so an energy independent fluctuation requires further study to determine systematically which components of $\nu_{dyn}$ are changing. UrQMD does not demonstrate a strong energy dependence, while another transport model (HSD, not shown) replicates the large increase in fluctuations at low energy observed by NA49. Both models treat resonance decay differently, which is one possibility to explain the different predictions. 

Figure \ref{ppi_excitation} shows the measured dynamical $p/\pi$ fluctuations as a function of incident energy, expressed as $\nu_{dyn,p/\pi}$. The measured fluctuations from the STAR experiment are plotted as the black stars, measurements from the NA49 experiment \cite{NA49_kpi_ppi} at the SPS as blue squares, and a transport model prediction from UrQMD (open red stars), using the STAR experimental acceptance. The published NA49 results have been converted to $\nu_{dyn}$ using the relation $\sigma_{dyn}^{2} \approx \nu_{dyn}$. STAR measures a general trend from larger, negative values of $p/\pi$ fluctuations at lower energies, which then increases towards zero at $\sqrt{s_{NN}}$ = 39 GeV. 
There is good agreement between the two different experiments at the lowest energies measured by STAR ($\sqrt{s_{NN}}$ = 7.7 and 11.5 GeV). The model prediction reproduce the qualitative trend shown in the data, but over predicts the magnitude at higher energies and crosses from negative to positive values. One contribution to this occurs as the models transition from associated to pair particle production at higher energies.

\section{Summary}

Results on dynamical particle ratio ($K/\pi$ and $p/\pi$) fluctuations from Au+Au collisions at $\sqrt{s_{NN}}$ = 7.7, 11.5, and 39 GeV have been presented. The dynamical fluctuations scaled by the measured charged particle multiplicity have been shown as a function of collision centrality. There is no apparent scaling with charged particle multiplicity for $K/\pi$ fluctuations in peripheral collisions at any energy, nor for $p/\pi$ fluctuations at $\sqrt{s_{NN}}$ = 11.5 and 39 GeV. However, multiplicity scaled $\nu_{dyn,p/\pi}$ at $\sqrt{s_{NN}}$ = 7.7 GeV is almost constant as a function of centrality. The STAR experiment has measured dynamical $K/\pi$ fluctuations that are approximately energy independent at these three energies and $p/\pi$ fluctuations that increase from larger negative values toward zero with increasing energy, starting from $\sqrt{s_{NN}}$ = 7.7 GeV. 

These are the first measurements of dynamical particle ratio fluctuations in Au+Au collisions below injection energy at RHIC. Additional energies will be analyzed as part of the QCD critical point search at RHIC, where STAR will be able to precisely measure dynamical $K/\pi$ and $p/\pi$ fluctuations, providing a comprehensive picture of these fluctuations from $\sqrt{s_{NN}}$ = 7.7-200 GeV. Dynamical fluctuations and their scaling properties will be studied for all separate charged particle ratio combinations.

\end{document}